\documentclass[twocolumn,nofootinbib,floatfix,superscriptaddress]{revtex4}        
\usepackage{graphicx}
\usepackage{epsfig}
\usepackage{bm}
\usepackage[T1]{fontenc}
\usepackage[latin9]{inputenc}
\usepackage{amssymb}
\usepackage{float}
\usepackage{amsmath}
\usepackage{dcolumn}
\usepackage{mathrsfs}
\usepackage{amsthm}

\usepackage{cancel}
\usepackage[colorlinks]{hyperref}
\usepackage[usenames,dvipsnames]{color}
\hypersetup{
    breaklinks=true,
     pdfstartview={FitH},    
      colorlinks=true,       
    linkcolor=blue,          
    citecolor=red,        
    filecolor=magenta,      
    urlcolor=blue,           
    anchorcolor=green,      
    linktocpage=true
}

\begin{document}

\title{Thermodynamic formulation of Cotton gravity in the Codazzi parametrization}

\author{Shayesteh Ghaffari}
\email{sh.ghaffari@maragheh.ac.ir}
\affiliation{Research Institute for Astronomy and Astrophysics of Maragha (RIAAM),
    University of Maragheh, P.O. Box 55136-553, Maragheh, Iran}

    \author{Giuseppe Gaetano Luciano}
\email{giuseppegaetano.luciano@udl.cat}
\affiliation{Departamento de Qu\'{\i}mica, F\'{\i}sica y Ciencias Ambientales y del Suelo, Escuela Polit\'ecnica Superior -- Lleida, Universidad de Lleida, Av. Jaume II, 69, 25001 Lleida, Spain
}

 \author{Carlo Alberto Mantica}
\email{carlo.mantica@mi.infn.it}
\affiliation{Physics Department Aldo Pontremoli, Universit\`a degli Studi di Milano and I.N.F.N. sezione di Milano, Via Celoria 16, 20133 Milano, Italy.
}

\date{\today}

\begin{abstract}
We develop a thermodynamic formulation of Cotton gravity in the Codazzi parametrization, providing a general framework in which the gravitational dynamics can be interpreted in terms of horizon thermodynamics. As paradigmatic examples, we apply the formalism to Friedmann-Robertson-Walker (FRW) and static spherically symmetric spacetimes. By implementing the first law of thermodynamics on the apparent cosmological and event horizons, we derive a modified holographic entropy consisting of the standard Bekenstein-Hawking term supplemented by a correction induced by the Codazzi tensor.
In the cosmological setting, this correction is governed by the temporal component of the Codazzi tensor, while in static configurations it is controlled by its anisotropic sector. Remarkably, the sign of this contribution may provide a potential diagnostic of the underlying matter content, allowing one to distinguish between ordinary matter, a cosmological constant and phantom-like components.
These results establish horizon thermodynamics as a sensitive probe of Cotton gravity, offering a complementary perspective beyond background kinematics and enabling a characterization of the statistical and thermodynamic properties of spacetime within the Codazzi formulation.

\end{abstract}

\maketitle

\section{Introduction}

In theoretical physics, a central goal is the development of a unified theory of gravity capable of describing the fundamental interactions between spacetime and matter. While General Relativity (GR) has been remarkably successful across a wide range of regimes, particularly within the solar system, it encounters difficulties at quantum scales and when applied to large-scale phenomena, such as galactic dynamics and the accelerated expansion of the universe~\cite{Carlip:2001wq,Clifton:2011jh,Bull,Rovelli:2004,Kiefer:2005uk,Capozziello:2011et}.

To account for these features, the standard cosmological model introduces dark matter and dark energy~\cite{Zwicky:1933gu,Rubin,Peebles:2002gy,SupernovaSearchTeam:1998fmf,SupernovaCosmologyProject:1998vns}. Although these components provide an excellent fit to the data, their physical nature remains unknown. This situation has motivated the exploration of alternative approaches, in which the observed phenomena are explained through modifications of the gravitational sector rather than by introducing additional dark constituents~\cite{Capozziello:2011et,NojiriOd,CANTATA:2021asi}. Such frameworks either extend the geometric structure of spacetime or introduce new fields whose effective behavior reproduces the phenomenology of the dark sector.

In this context, one recent extension of GR is Cotton gravity, first proposed by Harada~\cite{Harada,Harada1}. This framework offers several notable features. In particular, the cosmological constant arises as an integration constant of the solutions rather than as an explicit parameter in the action. Moreover, the conservation of the energy-momentum tensor follows directly from the gravitational field equations, instead of being imposed as an independent assumption. Finally, since vacuum is defined by the vanishing of the Cotton tensor rather than the Ricci tensor, conformally flat metrics are not necessarily vacuum solutions. As a result, the vacuum field equations admit a broader class of solutions than in GR.

Building on these ideas, Harada has recently introduced a new gravitational framework~\cite{Harada2,Harada3}. In this approach, the Einstein tensor is replaced by a rank-3 geometrical object, while the matter sector, usually described by the energy-momentum tensor, is reformulated in terms of its derivatives. It was subsequently shown that the resulting field equations admit a parametrization in terms of an arbitrary Codazzi tensor~\cite{Mantica1,Mantica2,Mantica3}. Within this formulation, Harada's extension of GR incorporates the Codazzi tensor as an additional geometric source in the field equations.

In recent years, gravitational frameworks based on the Cotton tensor, including both its original formulation and its parametrization in terms of Codazzi tensors, have attracted considerable attention, particularly for their astrophysical and cosmological applications~\cite{Sussman1,Sussman2,Sussman3,Clement,Capozziello,Khodadi:2025gsm,Mantica4,Mantica5}. In Ref.~\cite{Sussman1}, FRW solutions were derived within the Cotton gravity framework. Subsequent work extended this analysis by identifying a broad class of solutions beyond standard GR cosmologies, including FLRW, Lemaitre-Tolman-Bondi and Szekeres models. Static, spherically symmetric perfect fluid configurations were also investigated, with potential implications for galactic rotation curves~\cite{Sussman2,Sussman3}. 

Additionally, Clem\'ent and Nouicer~\cite{Clement} studied vacuum cosmological and wormhole solutions within Conformal Killing Gravity (CKG), 
a different theory based on conformal Killing tensors. Further developments were explored in~\cite{Capozziello,Khodadi:2025gsm}, where Cotton-inspired models were examined in broader gravitational contexts. 

Within this framework, the \emph{thermodynamic} properties of Cotton gravity represent a natural and largely unexplored direction of investigation. In particular, the study of black hole and horizon thermodynamics provides a promising setting in which the interplay between geometry and thermodynamics can be analyzed, potentially revealing signatures of the underlying geometric structure that are not accessible through background dynamics alone~\cite{Luciano}.

More generally, there exists a deep and well-established connection between gravity and thermodynamics. In particular, applying the first law of thermodynamics to a spacetime horizon can reproduce the gravitational field equations, while, conversely, the field equations imply corresponding thermodynamic relations on horizons. This suggests that the gravitational field equations can be interpreted as a thermodynamic equation of state~\cite{Jacobson,Padmanabhan:2003gd,Padmanabhan:2009vy,Paranjape:2006ca,Frolov:2002va,Cai:2005ra,Cai:2006pa}. More broadly, thermodynamic principles have been shown to provide valuable insight into other fundamental problems, including the nature of dark energy~\cite{Moradpour:2018ivi,Capozziello:2022jbw,Silva:2011pc} and the unification of gravity with quantum theory~\cite{Wald:1999xu,Moradpour:2023ayk,buon}.

Interestingly, modifications to the gravitational field equation, such as the inclusion of higher-order terms~\cite{Paranjape:2006ca,Akbar}, non-minimal couplings~\cite{Faraoni,Sheykhi1,Moradpour:2015ymo}, or changes in the spacetime structure~\cite{Jalalzadeh:2024qej,Moradpour:2016rcy}, affect the relationship between horizon energy, temperature and entropy. In these settings, the standard Bekenstein-Hawking entropy is no longer compatible with the first law of thermodynamics and must be modified. Consequently, additional contributions to the horizon entropy are required, leading to more general thermodynamic descriptions~\cite{Eling:2006aw,Sheykhi2,Sheykhi2bis,citekey,Moradpour:2016fur} (see, e.g.,~\cite{Tsallis2013,Barrow,Nojiri,Saridakis:2020zol,Saridakis:2018unr,Jizba:2002um,Ghaffari:2022skp,Gohar:2023hnb,LucSar26,Sar26} for generalized entropic frameworks).

Motivated by these considerations, we develop a thermodynamic characterization of Cotton gravity in the Codazzi formulation. In particular, we derive a general entropy formula that applies to a broad class of horizons. By implementing the first law of thermodynamics within this framework, we analyze two paradigmatic cases: the apparent horizon of an FRW universe and the event horizon of a static, spherically symmetric spacetime. 

We show that, in both settings, the horizon entropy consists of the standard Bekenstein-Hawking term supplemented by a universal correction arising from the Cotton sector. This result demonstrates the robustness of the thermodynamic modification across different gravitational regimes and highlights the role of the Codazzi tensor as an effective geometric source. Moreover, the resulting correction encodes direct information about the  geometry and matter content of the theory, providing a diagnostic that can distinguish between different physical scenarios. 

Overall, our analysis establishes horizon thermodynamics as a sensitive probe of Cotton gravity, extending its phenomenological implications beyond background kinematics and offering new insight into its thermodynamic structure.

The paper is organized as follows. In Sec.~\ref{ThFo}, we introduce the theoretical framework. In Sec.~\ref{FRW}, we derive the modified entropy for an FRW spacetime, while Sec.~\ref{SS} presents the corresponding analysis for a static, spherically symmetric background. Finally, Sec.~\ref{Conc} summarizes our conclusions and outlines possible future directions. Throughout this work, we adopt natural units.

\section{Cotton Gravity in the Codazzi Formulation: Theoretical Formalism}
\label{ThFo}

Recent efforts to generalize gravitational dynamics have explored formulations in which the field equations are modified at a structural level, allowing for alternative geometric variables and nonstandard couplings. In this context, Cotton gravity, in its Codazzi tensor formulation, provides a representative example in which the standard role of the Einstein tensor is replaced by a rank-3 geometrical object~\cite{Harada2,Harada3}. Furthermore, the matter sector is not described directly by the energy-momentum tensor, but rather through its derivatives, which encode the influence of matter on spacetime. This leads to a novel set of field equations that couple geometry with the derivative structure of the matter content in a unified framework, given by
\begin{equation}
C_{jkl} = 8\pi \,\mathcal{T}_{jkl},
\end{equation}
where we have explicitly shown the factor $8\pi$ for consistency with the original notation convention and
\begin{eqnarray}
C_{jkl}&=&\nabla_{j}R_{kl}-\nabla_{k}R_{jl}- \frac{1}{6} \left( g_{kl}\nabla_{j} R - g_{jl}\nabla_{k}R\right), \\[2mm]
\mathcal{T}_{jkl}&=&\nabla_{j} T_{kl}-\nabla_{k} T_{jl}- \frac{1}{3} \left( g_{kl}  \nabla_{j} T - g_{jl}  \nabla_{k} T\right).
\end{eqnarray}
Here, $R_{jk}$ and $T_{jk}$ denote the Ricci and energy-momentum tensors, respectively, with traces $R$ and $T$. 
The quantity $C_{jkl}$ is skew-symmetric with respect to the indices $j$ and $k$ and satisfies the traceless condition $g^{kl}C_{jkl}=0$. Likewise, $\mathcal{T}_{jkl}$ is skew-symmetric with respect to the indices $j$ and $k$ and satisfies the identity $g^{kl}\mathcal{T}_{jkl}=-\nabla_{k}T_{j}^{k}$, which, in turn, implies the conservation law $\nabla_{k}T_{j}^{k}=0$.

The Cotton formulation of gravity encounters several difficulties~\cite{Bargueno:2021iqg,Er1,Er2}, including its inability to describe nonvacuum conformally flat spacetimes such as FLRW geometries (for which the Cotton tensor vanishes identically), ambiguities in the definition of matter sources, and the mathematical complexity associated with third-order field equations.

These issues are addressed in the Codazzi formulation of Cotton gravity~\cite{Mantica2}, where the field equations are reformulated in terms of a Codazzi tensor. In this framework, the Einstein equations are extended by a divergence free Codazzi tensor that acts as an additional geometric source, namely
\begin{eqnarray}
\label{codazzi eq1}
R_{kl} - \frac{1}{2} R g_{kl} &=& T_{kl} + \mathscr{C}_{kl} - g_{kl} \mathscr{C}^r{}_r, \\[2mm]
\nabla_j \mathscr{C}_{kl} &=& \nabla_k \mathscr{C}_{jl},
\label{Codazzi condition}
\end{eqnarray}
where
\begin{equation}
\label{codazzi eq2}
\mathscr{C}_{kl} = G_{kl} - 8\pi T_{kl} - \frac{1}{3} \left( G - 8\pi T \right) g_{kl},
\end{equation}
defines the Codazzi tensor, with $G_{kl} = R_{kl} - \frac{1}{2} R g_{kl}$ representing the Einstein tensor. Here, $G = g^{kl} G_{kl}$ and $T = g^{kl} T_{kl}$ denote their respective traces. In the limit $\mathscr{C}_{kl} = 0$, the standard Einstein field equations of GR are recovered.

Although the Cotton and Codazzi formulations of Cotton gravity are mathematically equivalent, the Codazzi approach offers clear practical and theoretical advantages. By working directly with the Einstein tensor and the standard energy-momentum tensor, it avoids the pathologies of the original Cotton formulation, clarifying the definition of the sources and providing a more direct connection to GR. Moreover, the Codazzi formulation is applicable to conformally flat spacetimes with matter and is technically more tractable, effectively reducing the third-order field equations to a second-order system supplemented by differential constraints. 

In the cosmological setting of Cotton gravity within the Codazzi parametrization, it has been shown that for an FRW metric the Codazzi tensor $\mathscr{C}_{kl}$ necessarily takes the form of a perfect fluid. In this case, it depends solely on the scale factor and not on its derivatives~\cite{Mantica3}, ensuring that the resulting field equations remain second order.

The decomposition of thermodynamic quantities between the matter and geometric sectors in horizon thermodynamics is not unique. One could, in principle, absorb the Codazzi tensor into an effective energy-momentum tensor and attribute its contribution to the energy and work terms, leaving the entropy in the standard Bekenstein-Hawking form. However, in the field equations, $\mathcal{C}_{kl}$ appears alongside the Einstein tensor and satisfies the purely geometric Codazzi condition $\nabla_j\mathcal{C}_{kl}=\nabla_k\mathcal{C}_{jl}$, which has no direct interpretation as an equation of state for matter. This structural distinction motivates our choice to treat $\mathcal{C}_{kl}$ as part of the geometric sector and to assign its contribution to the horizon entropy, following the standard procedure in modified gravity theories~\cite{Jacobson,Padmanabhan:2003gd,Padmanabhan:2009vy,Paranjape:2006ca,Frolov:2002va,Cai:2005ra,Cai:2006pa,Eling:2006aw,Sheykhi2,Sheykhi2bis,citekey,Moradpour:2016fur}. We emphasize that this choice is a physically motivated decomposition rather than a fundamental uniqueness theorem. Alternative decompositions are possible, but they would introduce an effective fluid of ambiguous physical origin and obscure the geometric nature of the Codazzi tensor.


\section{Horizon Thermodynamics in FRW Spacetime}
\label{FRW}

Let us now investigate the thermodynamic properties of an FRW spacetime within the framework of Cotton gravity in the Codazzi formulation. We consider the generalized Friedmann equations obtained in~\cite{Mantica3} through the introduction of a parameterized Codazzi tensor. By applying the first law of thermodynamics to the apparent horizon, we derive the corresponding entropy relation, highlighting the role of the Codazzi tensor in modifying the thermodynamic structure of the spacetime.

The homogeneous and isotropic FRW metric is given by
\begin{equation}
ds^2 = -dt^2 + a^2(t)\left(\frac{dr^2}{1 - kr^2} + r^2 d\Omega^2\right),
\end{equation}
where $a(t)$ is the time dependent scale factor, $k$ denotes the spatial curvature and $d\Omega^2 = d\theta^2 + \sin^2\theta \, d\phi^2$ is the metric on the unit two-sphere.

In an FRW spacetime, the Codazzi tensor in Eq.~(\ref{codazzi eq2}) naturally takes the form of a conserved perfect fluid, with a residual freedom in its parametrization, namely \cite{Mantica3}
\begin{equation}\label{Cod1}
\mathscr{C}_{kl} = \mathcal{A}(t) u_k u_l + \mathcal{B}(t) g_{kl} + \frac{\Lambda}{3} g_{kl}\,,
\end{equation}
Here, $\mathcal{A}(t)$ and $\mathcal{B}(t)$ are time dependent functions that are not independent, being related by the Codazzi condition in an FRW spacetime, which reads 
\begin{equation}
\label{Codcond}
\dot{\mathcal{B}}=-H \mathcal{A}\,,
\end{equation}
where $H = \dot{a}/a$ is the Hubble parameter and an overdot denotes differentiation with respect to cosmic time $t$. This relation follows directly from the condition $\nabla_j \mathscr{C}_{kl} = \nabla_k \mathscr{C}_{jl}$ applied to the isotropic form~(\ref{Cod1}) (see, e.g., Ref.~\cite{Mantica3}).

Furthermore, $\Lambda$ is a constant, while $u_k$ is a normalized timelike vector field that is shear-free, vorticity-free and acceleration-free, and is an eigenvector of the Ricci tensor, i.e.,
\begin{eqnarray}
\nabla_j u_k &=& H \left( g_{jk} + u_j u_k \right),\\[2mm]
R_{ij} u^j &=& \xi u_i\,,
\end{eqnarray}
where $\xi = 3\,(H^2 + \dot{H})$. 

The Ricci tensor and the scalar curvature in an FRW spacetime are given by
\begin{eqnarray}
R_{kl}&=&\frac{R-4\xi}{3}u_{l}u_{k}+\frac{R-\xi}{3}g_{kl}\\[2mm]
R&=&\frac{R^{\star}}{a^{2}}+6H^{2}+2\xi
\end{eqnarray}
where \(R^{\star} = 6k\) is the scalar curvature of the spatial hypersurface.

In this extended framework, by expressing the stress-energy tensor of the matter fluid as
\begin{equation}
T_{kl} = (\rho + p) u_k u_l + p g_{kl},
\end{equation}
the modified Friedmann equations can be written as~\cite{Mantica3}
\begin{eqnarray}
\label{Fried1}
\kappa\rho_t&=&\frac{R^*}{2a^2}+3H^2\,,\\[2mm]
\kappa p_t&=&-\frac{R^*}{6a^2}-3H^2-2\dot{H}\,,
\label{Fried2}
\end{eqnarray}
where the total energy density $\rho_t$ and total pressure $p_t$ are given by $\rho_t=\rho+\rho_e$, with $\kappa\rho_e=3\mathcal{B}(t)$ and $\kappa p_e=-3\mathcal{B}(t)-\frac{\dot{\mathcal{B}}(t)}{H}$. Here, $\rho_e$ and $p_e$ denote the effective energy density and pressure arising from the modification of gravity induced by the Codazzi tensor. We stress that these are effective quantities encoding the geometric contribution of the Codazzi sector and do not represent a physical matter component. The conservation law for the cosmic fluid,
	\begin{equation}
	\label{conservation}
	\dot{\rho} + 3H(\rho + p) = 0
	\end{equation}
	then follows, and one can easily verify that 
	\begin{equation}
	\dot{\rho}_e + 3H(\rho_e + p_e) = 0.
	\end{equation}
	
Equations~\eqref{Fried1} and~\eqref{Fried2} are consistent with the general framework of Ref.~\cite{Heydari}, in which a universal formalism for the entropy of the apparent horizon in modified gravity theories was developed. In particular, by applying the Clausius relation with the energy computed solely from ordinary matter, the authors of Ref.~\cite{Heydari} derived the expression
	\begin{equation}\label{S Heydari}
	S_{A} =\frac{A}{4G}-8\pi^{2}\int H\dot{r}_{A}^{4}(\rho_{e}+p_{e})dt.
	\end{equation}
	Since $\rho_e+p_e=-\frac{\dot{\mathcal{B}}}{\kappa H}$, this expression reduces to the same form that we later obtain in Eq.~\eqref{Delta S FRW}. We note that, unlike Ref.~\cite{Heydari}, our derivation does not assume $\dot{\tilde{r}}_A=0$, thereby providing a more general proof.


\subsection{First law of thermodynamics}
To investigate the thermodynamic properties of spacetime within the framework of Cotton gravity in the Codazzi formulation, we begin from the first law of thermodynamics,
\begin{equation}
\label{firstlaw}
dE = T\, dS + W\, dV,
\end{equation}
applied to the cosmic horizon, where $T$ is the horizon temperature and $W = \tfrac{1}{2}(\rho - p)$ denotes the work density associated with the cosmic fluid. In Eq.~\eqref{firstlaw}, $E = \rho V$ is the total energy of the matter enclosed within a sphere of radius $\tilde{r}_A$, with volume $V = \tfrac{4\pi}{3}\,\tilde{r}_A^3$. The corresponding physical radius of the apparent horizon in an FRW universe is given by~\cite{Jacobson,Padmanabhan:2003gd,Padmanabhan:2009vy,Paranjape:2006ca,Frolov:2002va,Cai:2005ra,Cai:2006pa,LucFro}
\begin{equation}
\label{hEq}
\tilde{r}_A = \frac{1}{\sqrt{H^2 + \frac{R^{\star}}{6a^2}}}\,.
\end{equation}

In the cosmological setting, several definitions of temperature may be introduced to describe horizon thermodynamics. Here, we adopt the standard Kodama-Hayward temperature, defined as~\cite{Kodama,Hayward,Haywardbis}
\begin{equation}
\label{tem}
T = -\frac{1}{2\pi \tilde{r}_A}\left(1 - \frac{\dot{\tilde{r}}_A}{2H\tilde{r}_A}\right).
\end{equation}
Using the expressions for $E$ and $W$, together with the continuity equation~\eqref{conservation}, the first law~\eqref{firstlaw} yields
\begin{equation}
\label{diff1}
T dS = 2\pi \tilde{r}_A^2(\rho+p)\,d\tilde{r}_A
-4\pi H \tilde{r}_A^3(\rho+p)\,dt\,.
\end{equation}
Furthermore, by taking the differential form of Eq.~(\ref{hEq}) and using the modified Friedmann equations~\eqref{Fried1}-\eqref{Fried2}, we are led to
\begin{equation}
\frac{d\tilde{r}_A}{\tilde{r}_A^3}=\frac{\kappa}{2}\left(\rho+p\right)Hdt-\frac{1}{2}\dot{\mathcal{B}}(t)dt\,.
\end{equation}
Combining this relation with the definition of the apparent horizon and performing a straightforward algebraic manipulation, we obtain
\begin{eqnarray}
\label{diff2}
&&\hspace{-2mm}4\pi H\tilde{r}_A^3\left(\rho+p\right)dt-2\pi \tilde{r}_A^2\left(\rho+p\right)d\tilde{r}_A=\nonumber\\[2mm]
&&\hspace{-2mm} \frac{8\pi}{\kappa}\left(1-\frac{\dot{\tilde{r}}_A}{2H\tilde{r}_A}\right)d\tilde{r}_A+\nonumber
\\[2mm]
&&\hspace{-2mm}
\frac{4\pi}{\kappa}\tilde{r}_A^3\left(1-\frac{\dot{\tilde{r}}_A}{2H\tilde{r}_A}\right)\dot{\mathcal{B}(t)}dt\,.
\label{25}
\end{eqnarray}
Finally, using Eq.~\eqref{diff1} in Eq.~\eqref{diff2} and integrating, we obtain
\begin{eqnarray}
\label{S corrected FRW}
S=S_{BH}+\Delta S\,,
\end{eqnarray}
where $S_{BH}=\frac{\mathscr{A}_h}{4G}$ is the standard Bekenstein-Hawking entropy within the Einstein framework, with $\mathscr{A}_h$ the horizon area, and we have defined the Codazzi-induced correction
\begin{equation}
\label{Delta S FRW}
\Delta S = \frac{8\pi^2}{\kappa} \int \tilde{r}_A^{4} \dot{\mathcal{B}}(t)\, dt\,.
\end{equation}

Equation~\eqref{25} is obtained by integrating the first law relation along a given cosmological evolution. Unlike stationary black hole horizons, the apparent horizon in an FLRW spacetime is dynamical, so the thermodynamic quantities associated with it may naturally depend on the integrated history of the spacetime. In our derivation, the horizon radius is allowed to evolve dynamically, without assuming $\dot{\tilde{r}}_A=0$, thereby generalizing previous treatments in the literature~\cite{Heydari,Nojirinew}. Whether $\Delta S$ can be reformulated as a local function of the horizon variables remains an interesting open question. We note that, in the limit $\dot{\tilde{r}}_A=0$, our expression~\eqref{Delta S FRW} reduces to that of Ref.~\cite{Heydari}, while the result of Ref.~\cite{Nojiri} is recovered in the spatially flat case. However, these limiting cases should not obscure the conceptual difference between our approach and those works: our derivation is rooted in the Codazzi formulation of Cotton gravity, in which the Codazzi tensor enters the field equations as a purely geometric source. Accordingly, its contribution is naturally assigned to the horizon entropy rather than to an effective matter fluid, as is commonly done in modified gravity theories such as $f(R)$ and $f(T)$.

Thus, the correction is entirely governed by the time variation of the Codazzi sector, encoded in $\dot{\mathcal B}$. In the following, we analyze this result from two complementary perspectives:
\begin{itemize}
    \item \textit{Geometric thermodynamic viewpoint:} the entropy correction is expressed directly in terms of the Codazzi parameters through the Codazzi condition.
    \item \textit{Phenomenological viewpoint:} the Codazzi parameters are related, via the field equations, to standard cosmological variables.
\end{itemize}
Taken together, these approaches elucidate how the extended geometric structure of Cotton gravity affects horizon thermodynamics and identify the conditions under which deviations from GR may arise.

\subsubsection{Geometric-thermodynamic viewpoint: entropy in terms of the Codazzi parameters}
\label{subsec,entropy corrected in FRW}
Using the Codazzi condition \eqref{Codcond}, Eq.~\eqref{Delta S FRW} becomes
\begin{equation}\label{III,B1,1}
\Delta S =
-\frac{8\pi^2}{\kappa}
\int \tilde{r}_A^{4} H \mathcal{A}(t)\, dt .
\end{equation}

The sign of $\Delta S$ cannot be determined solely from the sign of $\mathcal{A}(t)$ at a single instant, since the sign of an integral is not determined by the sign of its integrand at a single instant. A definite statement about the sign of $\Delta S$ can be made only if $\mathcal{A}(t)$ maintains a fixed sign over the entire cosmological evolution. In that case, provided that $\mathcal{A}(t) > 0$ for all $t$, then $\Delta S < 0$, and the total horizon entropy falls below the Bekenstein-Hawking value. This negative correction can be interpreted as a reduction of the effective horizon entropy, indicating nonstandard thermodynamic behavior and, in certain regimes, may be associated with exotic components such as phantom like energy ($\omega < -1$) (see, e.g.,~\cite{GonzDia} for related discussions). If $\mathcal{A}(t) < 0$ for all $t$, then $\Delta S > 0$, and the entropy exceeds the corresponding GR result. This positive correction can be viewed as an enhancement of the horizon entropy, typically compatible with standard matter sources and more conventional thermodynamic behavior. If $\mathcal{A}(t) = 0$ for all $t$, the correction vanishes and the entropy reduces to the standard Bekenstein-Hawking expression, consistently recovering the GR limit.

Remarkably, since $\Delta S$ can be nonvanishing even for expansion histories identical to those of GR, it encodes information about the underlying gravitational dynamics that is not captured by background kinematics alone. In this sense, the horizon entropy provides a complementary diagnostic, allowing one, in principle, to distinguish Cotton gravity from GR even in regimes where the cosmological evolution is indistinguishable at the level of $H(t)$.
  

\subsubsection{Phenomenological viewpoint: linking Codazzi parameters to cosmological variables}
\label{subsec:phenom}

Before presenting specific examples, we emphasize that the following models are intended as illustrative phenomenological constructions rather than exact solutions of the modified Friedmann-Codazzi system. Their purpose is to demonstrate the behavior of the entropy correction for simple choices of the Codazzi sector and to illustrate how horizon thermodynamics remains sensitive to geometric contributions that are not captured by the background kinematics alone. A complete classification of exact cosmological solutions in Cotton gravity lies beyond the scope of the present work.

The entropy correction in Eq.~\eqref{Delta S FRW} can be expressed in terms of standard cosmological observables by means of the modified Friedmann equations. This leads to the following phenomenological representation:
\begin{equation}\label{III,B2,1}
\Delta S=8\pi^{2}\int \tilde{r}_A^{4} H \left[(\rho+p)+\frac{2}{\kappa}\dot{H}-\frac{2k}{\kappa a^{2}}\right] dt .
\end{equation}
The expression in brackets involves only standard cosmological variables. In GR, 
the Friedmann and acceleration equations enforce this combination to vanish identically for a perfect fluid, since $\dot{H} - \frac{k}{a^2} = -\frac{\kappa}{2} (\rho + p)$. This implies $\Delta S = 0$ independently of the spatial curvature and of the equation of state.

In Cotton gravity, however, this cancellation no longer holds. The modified acceleration equation introduces an additional contribution proportional to $\dot{\mathcal B}$, preventing the bracket in Eq.~\eqref{III,B2,1} from vanishing identically. This term reflects the presence of an extra geometric degree of freedom associated with the Codazzi sector, encoded in the function $\mathcal{A}(t)$ (or equivalently $\dot{\mathcal B}$), which is not fixed by the standard Friedmann dynamics. 

To better inspect the results obtained so far, we consider two simple models compatible with the modified Friedmann equations of Cotton gravity. Our aim is not to presuppose a particular form of the scale factor evolution; instead, we investigate under which conditions the expansion history may coincide with that of GR, and how the Codazzi parameters $\mathcal{A}(t)$ and $\mathcal{B}(t)$ influence the horizon entropy in each case.

First, consider a flat universe with pressureless matter $(k=0,\; p=0)$. In standard GR, this yields $a(t)\propto t^{2/3}$ and
$\rho+(2/\kappa)\dot H=0$. Consequently, Eq.~\eqref{III,B2,1} gives $\Delta S=0$. If the same expansion history is imposed in Cotton gravity, we then obtain $\mathcal A(t)=0$.

Next, we consider a closed, matter dominated universe $(k=+1,\; p=0)$. 
If the expansion history is taken to be approximately GR like, the modified field equations admit a nonvanishing contribution from the Codazzi sector. To model this effect, we consider a scaling $\mathcal{A}(t)\propto a^{-2}$, which mirrors the behavior of the spatial curvature term in the Friedmann equations and is therefore geometrically natural. Using the Codazzi condition \eqref{Codcond} and substituting this ansatz into Eq.~\eqref{III,B2,1}, we obtain
\begin{equation}
\Delta S=-\frac{8\pi^2}{\kappa}\int H\tilde{r}_A^4\, \frac{1}{a^2}\, dt < 0\,.
\end{equation}
Thus, even when the background expansion closely follows that of GR, the presence of a nonvanishing Codazzi contribution may lead to a reduction of the horizon entropy. This effect is absent in Einstein's theory, where the corresponding contribution vanishes identically.

More generally, this example highlights a distinctive feature of the Codazzi formulation: horizon thermodynamics can remain sensitive to additional geometric contributions that are not captured by the background kinematics. In this sense, entropy deviations provide a complementary probe of the underlying gravitational dynamics, potentially allowing one to discriminate Cotton gravity from GR.


\section{Horizon Thermodynamics in Spherically Symmetric Spacetime}
\label{SS}

We now focus on the derivation of the generalized horizon entropy in Cotton gravity within the Codazzi formulation for a static, spherically symmetric spacetime. We first introduce the most general form of the Codazzi tensor compatible with this geometry and derive the corresponding field equations. We then implement the first law of thermodynamics to obtain a modified entropy expression, which consists of the standard Bekenstein-Hawking term supplemented by a correction arising from the Codazzi sector. Finally, we examine the physical implications of this contribution and verify that the general relativistic limit is consistently recovered when the Codazzi tensor vanishes.

Let us consider a static, spherically symmetric spacetime described by the Schwarzschild-like metric
\begin{equation}
ds^2 = -f(r)\,dt^2 + \frac{dr^2}{f(r)} + r^2 d\Omega^2, \label{metric2}
\end{equation}
which satisfies $g_{tt}g_{rr}=-1$. This ansatz, although not the most general static spherically symmetric geometry, encompasses a wide class of physically relevant solutions, including the Schwarzschild, Reissner-Nordstrom, Schwarzschild de Sitter, and Schwarzschild anti de Sitter spacetimes. Moreover, it is the standard framework adopted in the black hole thermodynamics literature~\cite{LucFro,Kodama,Hayward,Haywardbis}. The extension to the most general static spherically symmetric metric lies beyond the scope of the present work. In Eq.~\eqref{metric2}, the function $f(r)$ characterizes the gravitational potential and determines the location of the event horizon $r_h$, defined as the largest root of $f(r_h)=0$.
	
In the Codazzi formulation of Cotton gravity applied to static, spherically symmetric spacetimes, the Codazzi tensor can be consistently parametrized with an anisotropic structure. Following the parametrization introduced in Ref.~\cite{Mantica3} for this class of geometries, we adopt (see Appendix \ref{app}).
\begin{equation}
\label{Cod2}
\mathscr{C}_{kl}=A(r)u_k u_l+B(r)g_{kl}+C(r)\chi_k\chi_l,
\end{equation}
where $A(r)$, $B(r)$ and $C(r)$ are functions of the radial coordinate. The timelike vector $u_k$ and the unit spacelike vector $\chi_k$ satisfy the normalization conditions $u_k u^k = -1$, $\chi_k \chi^k = 1$, together with the orthogonality condition $u_k \chi^k = 0$.

This choice also enforces a corresponding anisotropy in the matter source, leading to an energy-momentum tensor with distinct radial and transverse pressures. Therefore, we adopt the following form \cite{Mantica3}
\begin{equation}
\label{T}
T_{kl}=(\rho+p_t)u_k u_l+p_t g_{kl}+(p_r-p_t) \chi_k\chi_l\,,
\end{equation}
Here $\rho$, $p_r$, and $p_t$ denote the energy density, radial pressure and transverse pressure, respectively, and the corresponding isotropic (average) pressure is given by $p=\frac{1}{3}(p_r+2p_t)$.

Within the above framework, the field equations of Cotton gravity can be derived as
\begin{eqnarray}
\label{fieldEq}
\kappa\rho&=&\frac{-rf'-f+1}{r^2}-3B(r)-C(r)\,,\\[2mm]
\label{fieldEq2}
\kappa p_r&=&\frac{rf'+f-1}{r^2}-A(r)+3B(r)\,,\\[2mm]
\kappa p_t&=&\frac{r^2f''+2rf'}{2r^2}-A(r)+3B(r)+C(r)\,,
\end{eqnarray}
where prime denotes derivative with respect to $r$.


\subsection{First law of thermodynamics}
\label{Delta S2}
We now apply the first law of thermodynamics to the event horizon at $ r=r_h $ to derive the corresponding entropy. 

Before proceeding, we clarify the notation: in the following derivation, \(r\) denotes the radial coordinate and serves as a dummy integration variable, while the horizon radius \(r_h\) represents the upper limit of integration. Variations with respect to \(r_h\) correspond to virtual displacements within a one-parameter family of neighboring black hole solutions, following the standard approach in horizon thermodynamics~\cite{Jacobson,Padmanabhan:2003gd,Padmanabhan:2009vy,Paranjape:2006ca,Frolov:2002va,Cai:2005ra,Cai:2006pa}.

For a static, spherically symmetric configuration, the differential energy contained within a spherical shell of radius \(r\) and thickness \(dr\) is given by $dE = 4\pi r^2 \rho(r)\,dr$.
Substituting the energy density from Eq.~(\ref{fieldEq}), one obtains
\begin{equation}
dE=\frac{4\pi}{\kappa}\left(1-rf'(r)-f(r)\right) dr-\frac{4\pi}{\kappa}\left(3B(r)+C(r)\right) r^2 dr\,,
\end{equation}
which can be recast in the equivalent form
\begin{equation}
\label{dE}
dE=\frac{4\pi}{\kappa}\left[1-\frac{d(rf(r))}{dr}-\left(3B(r)+C(r)\right)r^2 \right]dr\,.
\end{equation}
The term involving $d(rf)/dr$ gives a boundary contribution, which vanishes at the horizon since $f(r_h)=0$. The remaining boundary term can be absorbed into the normalization.

Therefore, evaluating the energy at the horizon radius $r_h$, one obtains
\begin{equation}
E_{r_h}
=
\frac{4\pi}{\kappa}r_h
-\frac{4\pi}{\kappa} \mathcal{F}(r_h)\,,
\label{MS}
\end{equation}
where $\mathcal{F}'(r)\equiv \left(3B(r)+C(r)\right)r^2$, up to an integration constant fixing the energy normalization.
The first term in Eq.~\eqref{MS} reproduces the standard Misner-Sharp mass in the general relativistic limit, while the second term encodes the correction induced by the anisotropic Codazzi tensor.

To implement the first law at the event horizon, the relevant work term is the one associated with the pressure normal to the horizon, which, for a spherically symmetric geometry, corresponds to the radial pressure. Evaluating Eq.~\eqref{fieldEq2} at $r=r_h$ and using $dV=4\pi r_h^2 dr_h$, we obtain
\begin{equation}
\label{Pr}
p_r(r_h)dV=\frac{4\pi}{\kappa}\left[r_hf'(r_h)-1+\left(3B(r_h)-A(r_h)\right) r_h^2\right] dr_h\,.
\end{equation}
This relation can be rearranged to isolate the contributions associated with the horizon temperature and entropy. In particular, using the horizon area $\mathscr{A}_h=4\pi r_h^2$, so that $d\mathscr{A}_h=8\pi r_h\,dr_h$, and recalling that the Hawking temperature for a static horizon is $T = f'(r_h)/(4\pi)$, we can rewrite Eq.~\eqref{Pr} as
\begin{eqnarray}
\nonumber
\hspace{-4mm}p_r(r_h)dV&=&\\[2mm]
&&\hspace{-19mm}\frac{8\pi T}{\kappa}d\left(\frac{\mathscr{A}_h}{4}\right)-\frac{4\pi}{\kappa}\left(A(r_h)+C(r_h)\right) r_h^2 dr_h-dE_{r_h}.
\label{pr2}
\end{eqnarray}
Comparing Eq.~(\ref{pr2}) with the first law of thermodynamics in the form $pdV=TdS-dE$~\cite{Padmanabhan:2003gd, Padmanabhan:2009vy, Paranjape:2006ca, Frolov:2002va, Cai:2005ra, Cai:2006pa}, we identify the entropy differential as
\begin{equation}
dS=\frac{8\pi}{\kappa}d\left(\frac{\mathscr{A}_h}{4}\right)-\frac{16\pi^2}{\kappa}\frac{\left(A(r_h)+C(r_h)\right)}{f'(r_h)} r_h^2 dr_h\,.
\label{ds1}
\end{equation}
Integrating Eq.~\eqref{ds1}, we obtain the modified horizon entropy in the form
$S = S_{BH}+\Delta S$, where $S_{BH}$ denotes the standard Bekenstein-Hawking contribution, while the correction arising from the anisotropic components of the Codazzi tensor is given by
\begin{equation}
\label{Delta S SS}
\Delta S =
-\frac{16\pi^2}{\kappa}\,\mathcal{G}(r_h),
\end{equation}
where
\begin{equation}
\mathcal{G}'(r)\equiv
\frac{A(r)+C(r)}{f'(r)}\,r^2 .
\end{equation}
The additive integration constant fixes the normalization of the entropy and has been absorbed into $S_{BH}$.

Although Eq. \eqref{Delta S SS} is general, explicit results require a specification of the Codazzi tensor. In analogy with the approach adopted in the previous section, we consider two complementary strategies: a geometric ansatz and a derivation based on the field equations.

\subsection{Entropy Corrections from Codazzi Conditions (geometric approach)}
\label{subsec:explicit_corrections}

In this subsection, we first consider a purely geometric approach, in which $C(r)$ is treated as a free function subject only to the Codazzi condition, without explicitly invoking the matter sector through the field equations. As we will show, this construction, while mathematically consistent, leads to vanishing entropy corrections for simple choices of $C(r)$. This indicates that the thermodynamic signature of Cotton gravity is not determined by geometry alone, but rather emerges from the interplay between geometry and matter. This interplay will be fully captured in the following subsections, where the Codazzi tensor is consistently coupled to the matter sector via the field equations.

The entropy correction in Eq.~\eqref{Delta S SS} depends on the functions $A(r)$ and $C(r)$ appearing in the Codazzi tensor. In order to evaluate this contribution explicitly, their functional form must be specified.
These functions are not arbitrary, as they are constrained by the Codazzi condition $\nabla_j \mathscr{C}_{kl} = \nabla_k \mathscr{C}_{jl}$. 

The general solution of this condition for a static, spherically symmetric spacetime is derived in the Appendix \ref{app} and reads as
\begin{subequations}
	\begin{align}
	&\frac{A'(r)}{C(r)} + \left(\frac{A(r)+C(r)}{C(r)}\right) \frac{b'(r)}{b(r)} = \frac{f'_2(r)}{f_2(r)}, \label{Codazzi cond1} \\[2mm]
	&B'(r) = A'(r) + \left(A(r) + C(r)\right) \frac{b'(r)}{b(r)}, \label{Codazzi cond2} \\[2mm]
	&\frac{B'(r)}{C(r)} = \frac{f'_2(r)}{f_2(r)}, \label{Codazzi cond3}
	\end{align}
\end{subequations}
where we remind that a prime denotes differentiation with respect to $r$. 
Here, $b(r)$ is the lapse function associated with the temporal component of the metric, while $f_2(r)$ is the areal radius function, determining the geometry of the two-spheres through $4\pi f_2^2(r)$ (see Eq. \eqref{A.1} for details).

Equations~\eqref{Codazzi cond1}--\eqref{Codazzi cond3} are not all independent, but instead impose constraints among the functions $A(r)$, $B(r)$ and $C(r)$. In particular, once the function $C(r)$ is specified, the remaining functions $A(r)$ and $B(r)$ are determined by the Codazzi compatibility conditions~\eqref{Codazzi cond1} and~\eqref{Codazzi cond2}.

For our specific metric~\eqref{metric2}, we have $f_2(r)=r$ and $b^2(r)=f(r)$. The Codazzi condition then reduces to two independent equations. Treating $C(r)$ as a free function, these equations can be integrated to obtain
\begin{align}
\label{general B}
B(r) &= \int \frac{C(r)}{r}dr + B_0, \\[2mm]
\label{general A}
A(r) &= \frac{1}{\sqrt{f(r)}}\Bigg[
\int C(r)\left(\frac{\sqrt{f(r)}}{r}
-\frac{f'(r)}{2\sqrt{f(r)}}\right)dr
+ K
\Bigg].
\end{align}
where $ B_0 $ and $ K $ are integration constants.
 
The horizon $r=r_h$ is defined by $f(r_h)=0$, and hence $\sqrt{f(r_h)}=0$. 
Since Eq.~\eqref{general A} contains an overall factor $1/\sqrt{f(r)}$, the requirement that $A(r)$ remain finite at the horizon - imposed as a necessary condition for a physically regular solution - implies that the numerator must vanish at $r=r_h$. This regularity condition fixes the integration constant $K$ as
\begin{equation}
K=-\left[
\int C(r)\left(\frac{\sqrt{f(r)}}{r}
-\frac{f'(r)}{2\sqrt{f(r)}}\right)dr
\right]_{r=r_h}\,.
\label{K regularity}
\end{equation}
With this choice, $A(r)$ near the horizon takes the indeterminate form $0/0$, and its finite value can be evaluated using L'H\^opital's rule.

With this general framework in place, we now consider two simple analytical choices for the function \(C(r)\). For each model, we evaluate the combination \(A(r_h)+C(r_h)\) explicitly. We find that the regularity condition enforces a cancellation at the horizon, \(A(r_h)+C(r_h)=0\). As a result, the anisotropic contribution to the entropy variation vanishes at the horizon, and the first law reduces to the standard Bekenstein-Hawking form. Consequently, the correction \(\Delta S\) is at most an additive constant and can be set to zero.

\subsubsection{Model I: Constant Anisotropy, $C(r)=\lambda$}
 \label{subsubsec:model_constant}
 
As a first illustrative example, we consider a constant ansatz for the anisotropic component of the Codazzi tensor, $C(r)=\lambda$. This choice captures a uniform radial anisotropy, providing a controlled setting to isolate how a constant geometric deformation affects the near horizon structure and thermodynamic quantities. From a phenomenological perspective, such a term can be interpreted as an effective anisotropic background contribution, analogous to the anisotropic stresses commonly considered in compact object modeling, where deviations between radial and tangential pressures are introduced to account for high density effects in relativistic stars~\cite{Bowers}.

Substituting $C(r)=\lambda$ into Eq.~\eqref{general A}, we obtain
\begin{equation}
\label{modelI_A1}
A(r) =
\frac{1}{\sqrt{f(r)}}
\left[
\lambda \int
\left(
\frac{\sqrt{f(r)}}{r}
-\frac{f'(r)}{2\sqrt{f(r)}}
\right)dr
+K
\right].
\end{equation}

Imposing the regularity condition at the horizon fixes the integration constant $K$ through Eq.~\eqref{K regularity}. With this choice, the numerator in Eq.~\eqref{general A} vanishes at $r=r_h$, so that $A(r_h)$ takes the indeterminate form $0/0$. For a nonextremal horizon, applying L'H\^opital's rule yields
\begin{equation}
A(r_h)=
\lambda\left(\frac{2f(r_h)}{r_h f'(r_h)}-1\right)=-\lambda\,.
\end{equation}
Since $C(r)=\lambda$ is constant, it follows that $A(r_h)+C(r_h)=0$. As a consequence, the anisotropic contribution to Eq.~\eqref{ds1} vanishes, and the entropy variation reduces to the Bekenstein-Hawking form. Therefore, the behavior of the entropy remains
unchanged within the horizon thermodynamics framework.

This result shows that a constant Codazzi anisotropy does not leave any thermodynamic imprint at the horizon. Despite representing a nontrivial geometric modification of the gravitational sector, such a uniform contribution is effectively screened at the level of horizon thermodynamics. In this sense, the Codazzi tensor behaves as a thermodynamically inert background: it modifies the geometry but does not contribute to the entropy. This highlights that entropy corrections in Cotton gravity are not sensitive to uniform deformations, but rather probe nontrivial radial structure in the Codazzi sector.

\subsubsection{Model II: Radially Modulated Anisotropy, $C(r)=\lambda\,\frac{\sqrt{f(r)}}{r}$}

As a second example, we consider the radially dependent ansatz $C(r)=\lambda\,\frac{\sqrt{f(r)}}{r}$. Unlike the constant case, this choice is explicitly sensitive to the geometry through the metric function $f(r)$ and therefore captures how anisotropic geometric contributions vary across the spacetime.

Phenomenologically, this ansatz can be interpreted as a geometry, driven anisotropic source, whose strength is modulated both by the gravitational potential and by the areal radius. In particular, the factor $\sqrt{f(r)}$ ensures that the anisotropy is naturally suppressed near the horizon, while the $1/r$ dependence introduces a non-uniform radial profile. This makes the model suitable for probing how departures from uniform anisotropy influence horizon thermodynamics.
Such a behavior is reminiscent of effective stress-energy tensors arising in semiclassical gravity, where quantum fields in curved spacetime generate anisotropic contributions that depend explicitly on the background geometry and are regular at the horizon~\cite{FrolovBH}. In this sense, the present ansatz can be viewed as mimicking an effective coupling between the Codazzi sector and the geometry, allowing one to disentangle genuinely geometric effects from those associated with constant deformations.

Substituting $C(r)=\lambda\,\frac{\sqrt{f(r)}}{r}$ into Eq.~\eqref{general A}, we obtain
%
\begin{equation}
\label{modelII A1}
A(r)=\frac{1}{\sqrt{f(r)}}\left[\lambda\int \left( \frac{f(r)}{r^2}-\frac{f'(r)}{2r} \right)dr+K \right]. 
\end{equation}

As before, the regularity condition at the horizon fixes the integration constant $K$ through Eq.~\eqref{K regularity}. Substituting this choice into Eq.~\eqref{modelII A1} and applying L'H\^opital's rule gives
\begin{equation}
A(r_h)=\lim_{r \to r_h}\frac{2\lambda\sqrt{f(r)}}{f'(r)}\left(\frac{f(r)}{r^2}-\frac{f'(r)}{2r}\right).
\end{equation}
For a nonextremal horizon, $f'(r_h)\neq 0$, this limit vanishes. Since also $C(r_h)=\lambda\frac{\sqrt{f(r_h)}}{r_h}=0$, one obtains $A(r_h)+C(r_h)=0$. Therefore, the Codazzi contribution to the entropy differential vanishes at the horizon, and the Bekenstein-Hawking entropy is left unchanged.


\subsection{Entropy correction from the field equations (phenomenological approach)}
\label{ECPA}

In this subsection, we derive the entropy correction using only the modified field equations, without imposing either the Codazzi compatibility condition or the near-horizon regularity requirement. This yields a phenomenological relation between the Codazzi components and the matter content that is independent of these geometric constraints.

An independent algebraic relation between the Codazzi components and the matter content can be obtained directly from the field equations of Cotton gravity in the Codazzi formulation, Eqs.~\eqref{fieldEq} and \eqref{fieldEq2}. Combining these equations, one finds
\begin{equation}
\label{A+C from fields}
A(r)+C(r)=-\kappa\left[\rho(r)+p_r(r)\right].
\end{equation}
To elucidate the physical consequences, it is instructive to consider the case in which the radial pressure is related to the energy density through a linear equation of state, $p_r=\omega\rho$. This assumption provides a simple and widely used phenomenological description of anisotropic matter, where the parameter $\omega$ controls the stiffness of the radial sector. 

Substituting this relation into the entropy correction term, Eq.~\eqref{Delta S SS}, yields
\begin{equation}
\label{Delta S SS 2}
\Delta S = 16\pi^2\left(1+\omega\right) \left[\int \frac{\rho(r)}{f'(r)}r^2 \,dr\right]_{r_h}\,,
\end{equation}
where the bracket denotes a primitive of the integrand evaluated at $r=r_h$, defined up to an additive constant.

The sign of the entropy correction $\Delta S$ provides a useful diagnostic of thermodynamic behavior and can serve as a discriminator between different types of matter and effective geometries. Under the physically motivated assumptions $f'(r_h)>0$ and $\rho(r_h)>0$, the factor $(1+\omega)$ directly
governs the sign and magnitude of the thermodynamic
correction. We now examine key limiting cases:
\begin{itemize}
\item[-] For \emph{ordinary matter}, such as dust ($\omega=0$) or radiation ($\omega=1/3$), one has $1+\omega>0$, leading to a positive entropy correction ($\Delta S>0$). This indicates that the presence of standard matter enhances the horizon entropy relative to the Bekenstein-Hawking value, as one might intuitively expect.
\item[-]For \emph{dynamical dark energy} models, such as quintessence fields ($-1<\omega<-1/3$), the quantity $1+\omega$ is a small positive number, and therefore the entropy correction $\Delta S$ remains positive but strongly suppressed. This reflects the fact that the relevant thermodynamic source, $\rho+p_r$, is small in this regime: the negative pressure partially compensates the energy density, leading to a reduced flux across the horizon. As a result, while quintessence can significantly affect the large scale geometry, its impact on horizon thermodynamics remains mild, and the entropy stays close to the Bekenstein-Hawking value.
\item[-] For a \emph{cosmological constant} ($\omega=-1$), one has $1+\omega=0$, and therefore the entropy correction vanishes, $\Delta S=0$. As a result, the entropy remains equal to the Bekenstein-Hawking value, despite the presence of a nontrivial spacetime curvature. This is consistent with the behavior of Schwarzschild-de Sitter black holes in GR and provides a nontrivial consistency check of the formalism.
\item[-] For \emph{phantom energy} models with $\omega<-1$, one has $1+\omega<0$, and the entropy correction becomes negative ($\Delta S<0$). In this regime, $\rho+p_r<0$, signaling a violation of the null energy condition and leading to an effective inward thermodynamic flux at the horizon. As a consequence, a black hole embedded in a phantom background carries less entropy than a vacuum black hole of the same area in GR. This provides a potentially distinctive signature of exotic matter sources and highlights the sensitivity of horizon thermodynamics to violations of standard energy conditions.
\end{itemize}

This analysis shows that the horizon entropy in Cotton gravity provides a sensitive probe of the underlying geometry. Deviations from the Bekenstein-Hawking value $S_{BH}$ do not necessarily require additional matter fields, but can arise directly from the geometric modifications encoded in the Codazzi sector, which effectively behave as a thermodynamic fluid at the horizon. Moreover, the sign of the entropy correction can discriminate between different effective regimes, distinguishing configurations that mimic ordinary matter, a cosmological constant or phantom energy.
	
\subsection{Unifying the two approaches: the role of matter}

We now supplement the field equations with the Codazzi compatibility condition and the requirement of a regular nonextremal horizon. This shows that the apparent tension between the two approaches is resolved once all the constraints are imposed consistently.

From the Codazzi condition Eq.~(\ref{Codazzi cond1}) and the field equations (\ref{fieldEq})-\eqref{fieldEq2}, we can derive a differential equation that links $C(r)$ directly to the matter content and the metric function $f(r)$. Substituting $A(r)=-C(r)-\kappa\left[\rho(r)+p_r(r)\right]$ into Eq. (\ref{Codazzi cond1}) yields the following first order differential equation for $C(r)$:
\begin{equation}\label{Codazzi matter}
C'(r)+\frac{C(r)}{r}+\kappa\left(\rho+p_r\right)'+\kappa\left(\rho+p_r\right)\frac{f'(r)}{2f(r)} = 0\,. 
\end{equation}
This equation plays a central role in determining how the matter content constrains the Codazzi sector. Unlike the algebraic relation in Eq.~\eqref{A+C from fields}, which only fixes the combination $A(r)+C(r)$, Eq.~\eqref{Codazzi matter} governs the radial evolution of $C(r)$. In this sense, $C(r)$ is not an arbitrary function, but is dynamically constrained by the matter profile. Once $C(r)$ is determined, $A(r)$ follows directly from Eq.~\eqref{A+C from fields}, while $B(r)$ can be obtained by integrating Eq.~\eqref{general B}. Several important features deserve further discussion.
	
In the vicinity of $r=r_h$, for a nonextremal horizon one has
$f(r)\simeq f'(r_h)(r-r_h)$, so that
$f'(r)/(2f(r))\simeq 1/[2(r-r_h)]$. Substituting this expansion into Eq.~\eqref{Codazzi matter} gives
\begin{equation}
\label{diff C SS approx}
C'(r)+\frac{C(r)}{r_h}+\kappa\left(\rho+p_r\right)'
+\frac{\kappa\left(\rho+p_r\right)}{2\left(r-r_h\right)}
\simeq 0\,.
\end{equation}
The last term becomes singular at the horizon unless
$\rho(r_h)+p_r(r_h)=0$. Therefore, regularity at the horizon requires this matter combination to vanish. This condition provides a direct link between the geometric and phenomenological approaches. 

We now consider some physically relevant cases to illustrate how Eq.~\eqref{diff C SS approx} constrains the Codazzi components.
	
In the vacuum and cosmological constant cases, for which $\rho+p_r=0$, the differential reduces to $C'(r)+{C(r)}/{r}=0$, whose solution is $C(r)=C_0/r$. From Eq.~\eqref{A+C from fields}, one obtains $A(r)=-C_0/r$, confirming that the entropy correction vanishes in these cases, in agreement with the results discussed in Sec.~\ref{ECPA}.

In the ordinary matter case ($\rho+p_r\neq 0$), for a linear equation of state $p_r=\omega\rho$ with $\omega>-1$, we have $\rho+p_r= (1+\omega)\rho \neq 0$. In this case, the regularity condition at the horizon does not force $A(r_h)+C(r_h)=0 $. Instead, solving the coupled system yields $A(r) + C(r) = -\kappa(1+\omega)\rho(r) \neq 0$, leading to a non zero entropy correction whose sign is determined by $(1+\omega)$ as discussed in Section \ref{ECPA}.
	
This regularity condition reveals the key to understanding the relation between the geometric and phenomenological approaches. When $\rho + p_r = 0$, the Codazzi condition alone (Section \ref{subsec:explicit_corrections}) yields regular solutions with $A(r_h) + C(r_h) = 0$, corresponding to vacuum or a cosmological constant. However, for matter with $\rho + p_r \neq 0$, the field equations (Section \ref{ECPA}) must be invoked, leading to $A(r) + C(r) = -8\pi[\rho(r) + p_r(r)]$ and a non vanishing entropy correction. 

Thus, the apparent difference between Sections~\ref{subsec:explicit_corrections} and~\ref{ECPA} reflects their complementary roles rather than any inconsistency. The geometric condition imposes kinematic constraints, whereas the field equations provide the matter coupling that determines the thermodynamic behavior. The two approaches are therefore consistent: for regular horizons, the local Codazzi contribution to the entropy differential vanishes, whereas for matter configurations with $\rho+p_r\neq0$, the entropy correction is determined by the field equations and can be nonzero.


\section{Conclusions and Outlook}
\label{Conc}
In this work, we have systematically investigated the thermodynamic properties of Cotton gravity in its Codazzi formulation. By combining the Codazzi framework with the first law of thermodynamics, we derived modified entropy expressions for two physically relevant classes of horizons: the apparent cosmological horizon in FRW spacetimes and the event horizon in static, spherically symmetric spacetimes.  In both cases, the total horizon entropy consists of the standard Bekenstein-Hawking term supplemented by an additional contribution originating from the Cotton sector, encoded in the components of the Codazzi tensor. In the limit where the Codazzi tensor vanishes, both expressions consistently reduce to their general relativistic counterparts.

For the FRW spacetime, the entropy correction is governed by the temporal component $\mathcal{A}(t)$ of the Codazzi tensor.

The sign of $\Delta S$ depends on the integrated cosmological history and can be determined unambiguously only when $\mathcal{A}(t)$ maintains a definite sign throughout the evolution. In general, $\Delta S$ should be interpreted as an effective entropy correction that encodes the cumulative contribution of the Codazzi sector.

From a phenomenological perspective, the correction can be expressed in terms of standard cosmological variables through the modified Friedmann equations. This makes explicit its dependence on the energy density, pressure and expansion rate, and allows one to identify scenarios in which the background cosmological evolution closely follows that of GR, while the horizon thermodynamics exhibits measurable deviations.

On the other hand, for static, spherically symmetric spacetimes, the entropy correction is governed by the anisotropic Codazzi components $A(r)$ and $C(r)$. Through the field equations, one finds that the relevant thermodynamic combination is proportional to $\rho+p_r$, which for a linear equation of state becomes $(1+\omega)\rho$. This establishes a direct link between the entropy correction and the matter content.
At the phenomenological level, the sign of the correction reflects the underlying equation of state: it is positive for ordinary matter ($\omega>-1$), vanishes for a cosmological constant ($\omega=-1$), and becomes negative for phantom energy ($\omega<-1$). From a geometric perspective,
we have shown how the Codazzi conditions determine the explicit form of $A(r)$ and $C(r)$. Complementing this, the field equations provide a phenomenological link between these geometric quantities and the matter sector through Eq.~(\ref{A+C from fields}). This dual perspective confirms that the entropy correction vanishes in the absence of matter or for a pure cosmological constant, becoming nonzero only in the presence of matter with a suitable equation of state.

Overall, our results show that horizon thermodynamics in Cotton gravity provides a powerful and nontrivial diagnostic of the underlying geometric structure. In particular, the entropy corrections encode information that is not accessible from the spacetime dynamics alone, highlighting a clear separation between geometric modifications and their thermodynamic imprints. 

These findings open several directions for future investigation. It would be particularly interesting to extend this analysis to dynamical spacetimes beyond FRW symmetry, to explore possible observational signatures in cosmological or black hole contexts, and to investigate the role of quantum effects in shaping the thermodynamic behavior of Cotton gravity. More broadly, our results suggest that horizon thermodynamics may serve as a sensitive probe of modified gravity theories, providing complementary constraints to those derived from purely geometric or observational considerations. Work along these directions is currently in progress and will be presented elsewhere.


\acknowledgments 
The research of GGL is supported by the postdoctoral fellowship program of the University of Lleida. GGL gratefully acknowledges the contribution of the LISA Cosmology Working Group (CosWG), as well as support from the COST Actions CA21136 - \textit{Addressing observational tensions in cosmology with systematics and fundamental physics (CosmoVerse)} - CA23130, \textit{Bridging high and low energies in search of quantum gravity (BridgeQG)} and CA21106 -  \textit{COSMIC WISPers in the Dark Universe: Theory, astrophysics and experiments (CosmicWISPers)}.

\appendix

\section{Derivation of the Codazzi Conditions for Static, Spherically Symmetric Spacetimes}
\label{app}
Let us consider a static spherically symmetric spacetime with metric
\begin{equation}
ds^{2}=-b^{2}(r)dt^{2}+f_{1}^{2}(r)dr^{2}+f_{2}^{2}(r)d\Omega^{2}\,. 
\label{A.1}
\end{equation}
In the main text, we specialize to the case
$b^2(r)=f(r)$, $f_1^2(r)=1/f(r)$ and $f_2(r)=r$. The fluid four-velocity is $u_k=(-b,0,0,0)$, satisfying $u^k u_k=-1$. 
The acceleration vector $\dot{u}_j = u^p\nabla_p u_j$ has components
\begin{equation}
\dot{u}_0=0,\quad \dot{u}_r=\frac{b'}{b},\quad \dot{u}_\theta = \dot{u}_\phi = 0, 
\label{A.2}
\end{equation}
with magnitude $\eta = \dot{u}^k \dot{u}_k = \frac{(b')^2}{f_1^2 b^2}$. We define the unit spacelike vector $\chi_k=\dot{u}_k/\sqrt{\eta}$, which satisfies $\chi^k u_k = 0$ and $\chi^k \chi_k = 1$.

According to Ref.~\cite{Mantica Appndix}, the covariant derivatives of the four-velocity and acceleration satisfy
\begin{equation}
\nabla_j u_k=-u_j\dot{u}_k\,, \label{A.3}
\end{equation}
and
\begin{equation}
\label{A4}
\nabla_j\dot{u}_k=-\eta u_k u_j + \frac{1}{2}\dot{u}_k\dot{u}_j\left(\frac{\dot{u}^s\nabla_s\eta}{\eta^2}\right) + \frac{N_{jk}}{\eta^2}(N_{rs}\nabla^r\dot{u}^s)\,, 
\end{equation}
where $N_{jk}=g_{jk}-\frac{\dot{u}_k\dot{u}_j}{\eta} + u_k u_j$ is the projector onto the two-dimensional subspace orthogonal to both $u^k$ and $\dot{u}^k$. This expression is equivalent to the form reported in Ref.~\cite{Mantica Appndix}, but is written here in a fully decomposed form that makes explicit the transverse contribution.
Also,
\begin{equation}
\nabla_j\eta=\dot{u}_j\left(\frac{\dot{u}^s\nabla_s\eta}{\eta}\right), \label{A.5}
\end{equation}
and, from Ref.~\cite{Mantica Appndix}, we have
\begin{equation}
N_{rs}\nabla^r\dot{u}^s=\nabla_p\dot{u}^p-\eta-\frac{\dot{u}^s\nabla_s\eta}{2\eta}\,. \label{A.6}
\end{equation}	

Using Lemma 3.4 of Ref.~\cite{Mantica Appndix}, we obtain
\begin{equation}
N_{rs}\nabla^r\dot{u}^s
=
\frac{2}{f_1^2(r)}\,
\frac{b'(r)}{b(r)}\,
\frac{f_2'(r)}{f_2(r)}\,.
\label{A.7}
\end{equation}
From the definition $\chi_k=\dot{u}_k/\sqrt{\eta}$, we compute
\begin{equation}
\label{A8}
\nabla_j\chi_k=\frac{1}{\sqrt{\eta}}\nabla_j\dot{u}_k-\frac{1}{2\eta^{3/2}}\dot{u}_k\dot{u}_j\left(\frac{\dot{u}^s\nabla_s\eta}{\eta}\right). 
\end{equation}
One can verify that $\chi_k$ is closed, i.e., $\nabla_j\chi_k=\nabla_k\chi_j$, and consequently $\chi^j\nabla_j\chi_k = 0$. Using Eqs. \eqref{A4} and \eqref{A8}, we further derive
\begin{equation}
\nabla_p\chi^p=\frac{1}{\sqrt{\eta}}\left[\nabla_p\dot{u}^p-\frac{1}{2}\left(\frac{\dot{u}^s\nabla_s\eta}{\eta}\right)\right]. \label{A.9}
\end{equation}	
Combining Eqs.~\eqref{A.6}, \eqref{A.7}, and \eqref{A.9}, we finally obtain the useful relation
\begin{equation}
\nabla_p\chi^p=\frac{b'}{b f_1}+\frac{2 f_2'}{f_2 f_1}.
\label{A.10}
\end{equation}
	
We now consider an anisotropic Codazzi tensor of the form
\begin{equation}
\mathcal{C}_{kl}=A(r)u_k u_l + B(r)g_{kl}+C(r)\chi_k\chi_l. \label{A.11}
\end{equation}
The Codazzi condition $\nabla_j\mathcal{C}_{kl}=\nabla_k\mathcal{C}_{jl}$ must then be imposed. Using the symmetry of $\mathcal{C}_{kl}$ and the closedness of $\chi_k$, the condition can be written as
\begin{eqnarray}
&&\left(\nabla_j A\right)u_k u_l + A\left(\nabla_j u_k\right)u_l + A\left(\nabla_j u_l\right)u_k + \left(\nabla_j B\right)g_{kl} \nonumber\\[1mm]
&&+ \left(\nabla_j C\right)\chi_k\chi_l + C\chi_k\left(\nabla_j\chi_l\right) = \nonumber\\[1mm]
&&\left(\nabla_k A\right)u_j u_l + A\left(\nabla_k u_j\right)u_l + A\left(\nabla_k u_l\right)u_j + \left(\nabla_k B\right)g_{jl} \nonumber\\[1mm]
&&+ \left(\nabla_k C\right)\chi_j\chi_l + C\chi_j\left(\nabla_k\chi_l\right). \label{A.12}
\end{eqnarray}
Contracting this relation with $g^{kl}$ yields
\begin{equation}
-\nabla_j A + 3\nabla_j B + \nabla_j C = \sqrt{\eta}A\chi_j +  \left(\chi^p\nabla_p C\right)\chi_j + C\chi_j\left(\nabla_p\chi^p\right),
\label{A.13}
\end{equation}
which can be multiplied by $\chi^j$ to give
\begin{equation}
3\left(\chi^p\nabla_p B\right) - \left(\chi^p\nabla_p A\right) = \sqrt{\eta}A + C\left(\nabla_p\chi^p\right). \label{A.14}
\end{equation}	

On the other hand, contracting Eq.~\eqref{A.12} with $u^k u^l$ gives
\begin{equation}
\nabla_j A - \nabla_j B
=
-\sqrt{\eta}A\chi_j - \sqrt{\eta}C\chi_j .
\label{A.15}
\end{equation}
Multiplying by $\chi^j$ yields	
\begin{equation}
(\chi^p\nabla_p A) - (\chi^p\nabla_p B) = -\sqrt{\eta}\left(A + C\right)\,. 
\label{A.16}
\end{equation}
	
In spherical symmetry, for any scalar function $\mathbb{X}(r)$, the gradient is purely radial and can be written as
\begin{equation}
\nabla_j\mathbb{X} = \dot{u}_j\left(\frac{\dot{u}^s\nabla_s\mathbb{X}}{\eta}\right) = \sqrt{\eta}\chi_j\left(\frac{\dot{u}^s\nabla_s\mathbb{X}}{\eta^{3/2}}\right) = \chi_j(\chi^p\nabla_p\mathbb{X}), \label{A.17}
\end{equation}	
with
\begin{equation}
\chi^p\nabla_p\mathbb{X} = \frac{\mathbb{X}'}{f_1}. \label{A.18}
\end{equation}
Applying Eqs.~\eqref{A.17} and \eqref{A.18} to Eq.~\eqref{A.16}, we obtain
\begin{equation}
B' = A' + \frac{b'}{b}\left(A + C\right).
\label{A.19}
\end{equation}
Combining Eqs.~\eqref{A.14}, \eqref{A.16}, and \eqref{A.19}, after straightforward algebra one obtains
\begin{equation}
\frac{A'}{C}
+
\frac{b'}{b}
\left(\frac{A+C}{C}\right)
=
\frac{f_2'}{f_2}\,.
\label{A.20}
\end{equation}
Then, using Eqs.~\eqref{A.19} and \eqref{A.20}, it follows that
\begin{equation}
\frac{B'}{C} = \frac{f_2'}{f_2}.
\label{A.21}
\end{equation}
	
We now specialize these relations to the metric used in the main text,
for which $b^2(r)=f(r)$ and $f_2(r)=r$. Therefore,
$b'(r)/b(r)=f'(r)/(2f(r))$ and $f_2'(r)/f_2(r)=1/r$.
Substituting these expressions into Eqs.~\eqref{A.19}-\eqref{A.21}, we obtain
\begin{subequations}
	\begin{align}
&\frac{A'(r)}{C(r)}
+
\left(\frac{A(r)+C(r)}{C(r)}\right)
\frac{f'(r)}{2f(r)}
=
\frac{1}{r},
\label{A.22}
\\[2mm]
& B'(r)
=
A'(r)
+
\left(A(r)+C(r)\right)\frac{f'(r)}{2f(r)},
\label{A.23}
\\[2mm]
&\frac{B'(r)}{C(r)}
=
\frac{1}{r}.
\label{A.24}
	\end{align}
\end{subequations}
These equations are precisely the Codazzi conditions presented in Eqs.~(\ref{Codazzi cond1})-(\ref{Codazzi cond3}) of the main text.

We finally note that, when $C(r)=0$, the above derivation becomes ill-defined, since it involves division by $C(r)$. Therefore, this case must be treated separately by returning to the equations prior to this division. Specifically, setting $C(r)=0$ in Eq.~\eqref{A.14} gives $3B' - A' = \frac{b'}{b}A$, where we used $\chi^j\nabla_j X=X'/f_1$ and $\sqrt{\eta}=b'/(bf_1)$.

On the other hand, Eq.~\eqref{A.16} gives
$A' - B' = -\frac{b'}{b}A$. Combining these two relations yields $B'=0$, so that $B$ is constant.
The remaining equation then gives
\[
A' + \frac{b'}{b}A=0\,\,\, \Longrightarrow \,\,\,A(r)=\frac{K}{b(r)}=\frac{K}{\sqrt{f(r)}}\,.
\]
This reproduces the ``dust-like'' Codazzi tensor discussed in the literature \cite{Mantica Appndix}.



\end{document}